\documentclass[11pt]{article}
\usepackage{latexsym} 
\usepackage{graphics}

\newcommand{\ie}{{\it i.e. }}
\newcommand{\eg}{{\it e.g. }}
\newcommand{\Z}{{\sf Z\hspace{-1ex}Z}}        
\newcommand{\R}{\mbox{\rm I}\hspace{-.5ex}\mbox{\rm R}}
\newcommand{\be}{\begin{equation}}
\newcommand{\ee}{\end{equation}}
\newcommand{\rep}[1]{\sigma^{(#1)}}
\newcommand{\ba}{\begin{array}}
\newcommand{\ea}{\end{array}}  
\newcommand{\bea}{\begin{eqnarray}}
\newcommand{\eea}{\end{eqnarray}}  
\newcommand{\dlangle}{\langle\hspace{-.6ex}\langle}
\newcommand{\drangle}{\rangle\hspace{-.6ex}\rangle}
\newcommand{\sign}{\mathrm{sign}}
\newcommand{\vol}{\textbf}

\newcommand{\plaq}{\Box} 

\newcommand{\email}{e-mail:\ }

\begin{document}
\title{\bf Renormalization Group Approach to Spin Glass Systems}
\author{G. Parisi $\mbox{}^a$ 
\and R. Petronzio $\mbox{}^b$
\and F. Rosati $\mbox{}^c$ }
\date{November 17, 2000}
\maketitle
\begin{center}
{\it a)} Dipartimento di Fisica and INFN, 
Universit\`a di Roma ``La Sapienza'',\\
P.le Aldo Moro 2, I-00185 Roma, Italia\\
\email{giorgio.parisi@roma1.infn.it}\\
{\it b)} Dipartimento di Fisica and INFN,
Universit\`a di Roma ``Tor Vergata'',\\
via della Ricerca Scientifica 1, I-00133 Roma, Italia\\
\email{roberto.petronzio@roma2.infn.it}\\
{\it c)} Dipartimento di Fisica,
Universit\`a di Roma ``Tor Vergata'',\\
via della Ricerca Scientifica 1, I-00133 Roma, Italia\\
\email{francesco.rosati@roma2.infn.it}\\
\end{center}
\bigskip

\begin{abstract} 
A renormalization group transformation suitable for spin glass models
and, more generally, for disordered models,
is presented. The procedure is non-standard in both the nature of
the additional interactions and the coarse graining transformation,
that is performed on the overlap probability measure (which is clearly
non-Gibbsian).  Universality
classes are thus naturally defined on a large set of models, going from
$\Z_2$ and Gaussian spin glasses to Ising and fully frustrated models,
and others. 

The proposed analysis is tested numerically on the
$\Z_2$ Edwards--Anderson model in $d=4$. Good estimates of the
critical index $\nu$ and of  $T_c$ are obtained, and an RG flow diagram is 
sketched for the first time.


\end{abstract}

\bigskip

In the understanding of critical phenomena a crucial role is played by
the renormalization group (RG) analysis~\cite{Wilson71, DicastroJona}. 
It provides a
theoretical foundation of the universality principle and of the
scaling theory, as well as a method for a direct evaluation of critical
indices. 
Despite the fact that both universality and scaling are largely 
used in the study of
spin glass (SG) models, the RG approach has not been fully
developed~(see however \cite{DeDominicis:1997, Palassini:1999}). 
Classical
tools of RG analysis are not suitable for spin glasses, mainly because
a direct transformation of the Hamiltonian is impossible. The random
nature of the couplings require also the additional interactions to be
introduced as random variables. But averaging over disorder makes it
impossible to consider simple Boltzmann--Gibbs distributions, and to
use the spin variables for a coarse graining transformation.

Therefore we propose a RG transformation on the overlap distribution,
in the spirit of the probabilistic interpretation of the
renormalization group~\cite{DicastroJona}. 
Universality
classes are naturally defined on a large set of models, going from
$\Z_2$ and Gaussian spin glasses to Ising and fully frustrated models,
and others.


\section{The model}
A large part of the discussion will be general, but to be explicit let us
consider the Edwards--Anderson spin glass model~\cite{EA75}. 
Let $\sigma_x$ be
Ising spins located at the sites of a $d$-dimensional cubic lattice 
($x\in \Lambda\subset\Z^d$).  
The Hamiltonian of the model is given by
\be
H(J,\,\sigma) = - \sum_{\langle x,\,y\rangle} J_{xy}\sigma_x\sigma_y
,\label{hamiltonian}
\ee
where the sum is over the couples of nearest neighbouring sites.
Some kind of boundary conditions (\eg periodic) are also provided. 
The quenched disordered interactions
$J_{xy}$'s are independent random variables with zero mean and unit variance.
Denoting by $E$ expectation on $J$ variables, we have
\be 
E(J_{xy}) = 0,\;\;E(J_{xy}^2) = 1
.\ee
A Boltzmann--Gibbs measure on the spin variables is introduced, following
the usual rules of statistical mechanics. It will be denoted by
angular brackets $\langle\cdot\rangle$. Expectation on the disorder
is taken only after Boltzmann averages are calculated,
and the thermodynamic limit for the appropriate quantities 
is eventually taken afterwards. 
The model is symmetric under the {\it gauge} transformation
defined by
\be
\left\{ \ba{l}
 J_{xy}\rightarrow J'_{xy} = \varepsilon_x \varepsilon_y J_{xy} \\
 \sigma_x\rightarrow \sigma'_x = \varepsilon_x\sigma_x \\
\ea \right.
\ee
where $\varepsilon_x = \pm 1$ are the gauge group parameters.
Let us introduce the overlap variables. Consider $s$ replicas
(copies) of the spin variables $\rep{a},\, a=1,\ldots,s$. 
The Hamiltonian of the replicated system is given by 
${\mathcal H} = \sum_a H(J,\,\rep{a})$, therefore replicas are
independent from each other, but feel the {\it same} disorder
configuration $J$. For any couple of replicas $(a,b)$, introduce 
the {\it site overlap} $q^{(a,b)}_x$ 
\be
q^{(a,b)}_x = \rep{a}_x\rep{b}_x\,\in \Z_2,
\ee
The overlap probability distribution $\mu$ can be implicitly defined 
through the overlap expectations, that involve both the thermal
average and the average $E$ over disorder. For any smooth function $F$
define 
\be
\dlangle F\left( q^{(12)}_x, q^{(23)}_y,\ldots\right) \drangle
= E\left( \langle F\left( q^{(12)}_x,q^{(23)}_y,\ldots\right)
\rangle\right),
\label{overlap-av}
\ee
where expectation with respect to the $\mu$
distribution is denoted by $\dlangle\cdot\drangle$.  
All physical observables can be expressed in terms of overlap
observables, so that the full physical meaning of these models is
contained in the overlap probability measure.

Let us introduce the average over the volume of the site overlap,
the total overlap, often simply referred to as ``overlap'', 
\be
q^{(a,b)} = \frac{1}{|\Lambda|} \sum_x q^{(a,b)}_x
.\ee
Given 2 replicas, the distribution of the overlap $q^{(1,2)}$ will
be denoted by $P^{(2)}$. 
An other interesting observable is the correlation function of site
overlap,
\be
C(r)=\frac{1}{|\Lambda|} \sum_x\dlangle
q_x^{(1,2)}q_{x+r}^{(1,2)}\drangle_c
.
\ee


\section{RG \& SG}

Two main difficulties arise in the application of RG analysis to
spin glasses: the choice of the coarse graining transformation and a
correct parameterization of the space of Boltzmann measures -- that is,
the nature of the additional interactions. 
Let us begin by the second. 
In spin glasses the single interactions 
are not set explicitly, but only through their distribution. 
Instead of coupling constants we have distributions of couplings, and
different distributions give rise to different physical models.

Therefore, our proposal is 
to keep the form of the Hamiltonian fixed, but
to take into account a large space of disorder distributions, such
that the $J$'s are not independent variables. 
This choice is coherent with 
the random nature of the couplings, and introduces
effective long range interactions between the spins
through the correlations of the $J$'s. 
More generally, new explicit spin
interactions could be added, carried by new disorder variables.
This procedure would lead to a very large space of models that can
be very interesting to be studied, but in our opinion it is unnecessary
for our present purpose. 
Notice that it would be redundant to introduce new spin interactions
depending only on the old disorder variables.

The disorder distribution must respect gauge invariance,
therefore it will be parameterized as follows
\be
\rho(J) = \exp\left( {\sum}_i K_i \, W_i \right)
,\ee
where the $K_i\in\R$ are parameters and the $W_i$ are the Wilson's
loops, \ie  products of $J$'s along a closed path. 
Therefore we can
consider the following general disorder probability distribution  
\be
\rho_K = C_K\,\exp\left( 
K_1\sum_{\langle x,\,y\rangle} J_{xy}^2
+ K_2\sum_{\langle x,\,y\rangle} J_{xy}^4
+ K_3 \sum_\alpha \plaq_\alpha  
+ o(J^4) \right)
,\ee
where $C_K\in\R$ is a normalization constant, and the symbol $\plaq$ 
denotes the
plaquette terms of the kind $J_{x,y} J_{y,z} J_{z,w} J_{w,x}$. 
Expectation with respect to this distribution will be
denoted by $E_K$. The Gaussian Edwards--Anderson model
corresponds to $K_1 = -1/2,\ K_i=0\ \forall i\not=1$, 
while the $\Z_2$, E--A model is obtained in the limit
\bea
K_1, K_2 &\rightarrow& \infty,\ \mathrm{s.t.}\ K_1/K_2 = -2, \\
K_i &=& 0 \;\;\;\forall i>2\;.\nonumber
\eea
Another interesting distribution is obtained from the $\Z_2$ E--A model,
adding a plaquette term: 
\bea
K_1, K_2 &\rightarrow& \infty,\ \mathrm{s.t.}\ K_1/K_2 = -2, \nonumber\\
K_3 &\not =& 0, \label{pure_gauge}\\
K_i &=& 0 \;\;\;\forall i>3\;.\nonumber
\eea
Considering the disorder variables only, 
this is the well known pure gauge $\Z_2$ model~\cite{Balian:1975ir}.  
In the limit $K_3\rightarrow \infty$, frustration disappears and the 
Ising model is obtained, up to a simple gauge transformation.
Attention is to be payed to ensure that all Polyakov loops are
positive, to avoid the appearance of interfaces. 
A change of behaviour is to
be expected at the deconfinement transition. In dimension $d=4$, the
transition is of the first order: at
$|K_3|\geq K_3^c=log(\sqrt{2}+1)/2\simeq 0.4407$ the average plaquette jumps to
$|E_K(\plaq)|\sim 1$. The opposite limit, $K_3\rightarrow -\infty$,
leads to the fully frustrated model.

According to eq.~(\ref{overlap-av}), we introduce 
the overlap probability distribution $\mu_K$ corresponding to the
disorder measure $E_K$. Let us denote by $\mathcal{M}$ the space of 
such distributions. 
Notice that the dependence on $\beta=T^{-1}$ of $\mu_K$ could be 
included in the $K$'s
parameters, performing the simple substitution $J\rightarrow TJ$, so that the 
Boltzmann factor does
not contain $\beta$, and $E_\beta(J^2)=\beta^2$. 

The choice of the coarse graining transformation is related to the
nature of the order parameter. Block transformations of the spins are
suitable for the Ising model because the order parameter is the
average magnetisation, and the diverging correlation length is given
by the spin correlation function.
For spin glasses the elementary physical
observable is the site overlap $q^{(ab)}_x$, while
the phase transition is characterized by a functional order
parameter, the distribution $P^{(2)}$ of the overlap between two
replicas. 
This distribution is non-trivial in the broken replica symmetry
phase~\cite{MPV, Parisi_SG_rev00}.
The correlation length defined by the correlation function $C(r)$
diverges at the critical point.
For those reasons, we will perform a block transformation on the site 
overlap, leading to a renormalized overlap defined by
\be
q_B = \sign\left( {\sum}_{x\in B} q_x \right)
.\ee
The RG transformation $R_l$ for a rescaling factor $l$ 
is defined on the overlap distribution $\mu$ as follows:  
\bea
&& \mu^{(l)}(q') = \sum_q \hat{R}_l(q',\,q) \, \mu(q), \\
&& \hat{R}_l(q',\,q) = \prod_{x'}
\delta\left(q'_{x'},\, \sign\left( {\sum}_{x\in B_l(lx')} q_x \right) \right),
\eea
where $x'=x/l$ is the rescaled lattice coordinate and $B_l(lx')$ 
is the block of side $l$ located in $x=lx'$.
Let us assume the following\\
{\it\bf{Ansatz}}  The set $\mathcal{M}$ of all the overlap 
distributions $\mu_K$ that can be represented through the parameters $K$
is invariant under the RG transformations, \ie 
\be
R_l \left( \mu_{K^{(1)}} \right) = \mu_{K^{(l)}}\in\mathcal{M}.
\ee
Exploiting the Ansatz, we can introduce the RG transformation on the
parameters space
\be
K^{(l)} = R_l K^{(1)},
\ee
and look for the fixed points $K^\ast = R_l K^\ast$ and the corresponding
universality classes. The proposition that Gaussian and $\Z_2$ SG
models belong to the same universality class is therefore formulated in a
precise theoretical framework. 
We expect to find the fixed point of the spin glass
transition in the deconfined phase, while for $|K_3|> K_3^c$ the phase
transition is to be characterized by different fixed points, somehow
related to the Ising model (or, more likely, to the diluted Ising
model) for $K_3$ positive, and to the fully frustrated model for 
$K_3< -K_3^c$   
and of the Ising model ferromagnetic transition. 
The relations with the fully frustrated model may be understood,
and the intermediate cases would also be clarified, in terms of
universality classes.

We have so far introduced a parameterization of the space of overlap
probability measures and a renormalization group transformation in
this space, suitable for spin glass models. The parameterization is rather
implicit, and not easy to be employed in analytic calculations. 
Therefore, this RG transformation has been tested numerically, with the
Monte Carlo method. 


\section{Finite-size renormalization group}

The RG transformation defined in the previous section has been applied
to the 4-dimensional E--A spin glass model with periodic b.c., 
using Monte Carlo method. 
A typical problem of this approach is that one is limited
by the finite size of the lattice to a few RG iterations. Moreover, it is
practically very difficult to consider a large number of
parameters, giving rise to truncation errors.
To minimize the effect of these approximations, the following method 
has been applied, as introduced in~\cite{BP89}. 
For a lattice of linear size $L$, block variables of size $L/2$ are
defined, obtaining a $2^d$ system after one RG iteration. Therefore
the number of possible $K$ parameters in the renormalized system is
very limited and the truncation error is controllable. 
By varying the initial lattice size $L$ we obtain 
many sets of parameters $K'(K;\,L)$, renormalized by different
rescaling factors $l=L/2$, but realized in the same effective lattice $2^d$. 
Therefore are avoided the systematic errors due to the comparison of
$K$'s acting in different effective lattices and, thus, with different
truncations. 
At the critical point, the couplings are independent 
of the rescaling factor, and the $K'(K_c,\,L)$, for different
$L$'s, will coincide. This gives an estimate of $K_c$, \ie of $T_c$.
For any pair of sizes a different estimate $K_c(L_1,\,L_2)$ is
obtained, with a small dependence on $L_1,\,L_2$ vanishing for large
sizes. 
An estimate of the thermal critical index $\nu$ can be obtained
independently from each coupling, according to the well known formula
\be
\nu^{-1} = y_T(K_i) = \frac{  \ln \left( \frac{dK_i}{d\beta }(L_1) /
	\frac{dK_i}{d\beta }(L_2) \right) } { \ln (L_1/L_2) }
,\label{critical_index}
\ee
Clearly, the numerical values of the resulting fixed point $K^\ast$
are not the true infinite volume values. However, the estimates of
$K_c$ and of the critical index are weakly affected by finite-size
effects.
  
The general idea of interpreting the ratio of different lattice sizes
as RG scaling factor was developed by Nightingale and is known as
phenomenological~RG~\cite{Nightingale}. 
In this approach the test quantities are the
specific heat or the susceptibility of the original system and not
renormalized quantities like in the method exploited in this paper.
Moreover, it does not contain any effective Hamiltonian (or effective
measure).

In the Monte Carlo runs we have considered the $\Z_2$ E--A model in 4
dimensions, with periodic boundary conditions. Only one additional
parameter, $K_3$, has been introduced in the effective measure.
The RG transformation used was a majority rule on the block overlap, with a
tie-breaker if zero.
The renormalized parameters where obtained with a matching condition
on the overlap probability distribution.
The complete procedure is as follows.
\begin{enumerate}
\item 
MC run on $L=2$ system, for a wide range of temperatures and
values of the $K_3$ parameter. The configurations of the disorder $J$
where obtained by independent pure-gauge MC runs.
\item
MC run on large systems, $L=6,8,10,12$, measuring the renormalized
overlap distribution, for fixed $K_3=0$ and various temperatures.
\item
The renormalized overlap distribution 
is matched to the overlap distribution measured in the $L=2$
system, finding the effective parameters $T',\ K_3'$.
\end{enumerate}

\section{Numerical results}

\begin{figure}
\resizebox{\textwidth}{!}{\rotatebox{270}{{\includegraphics{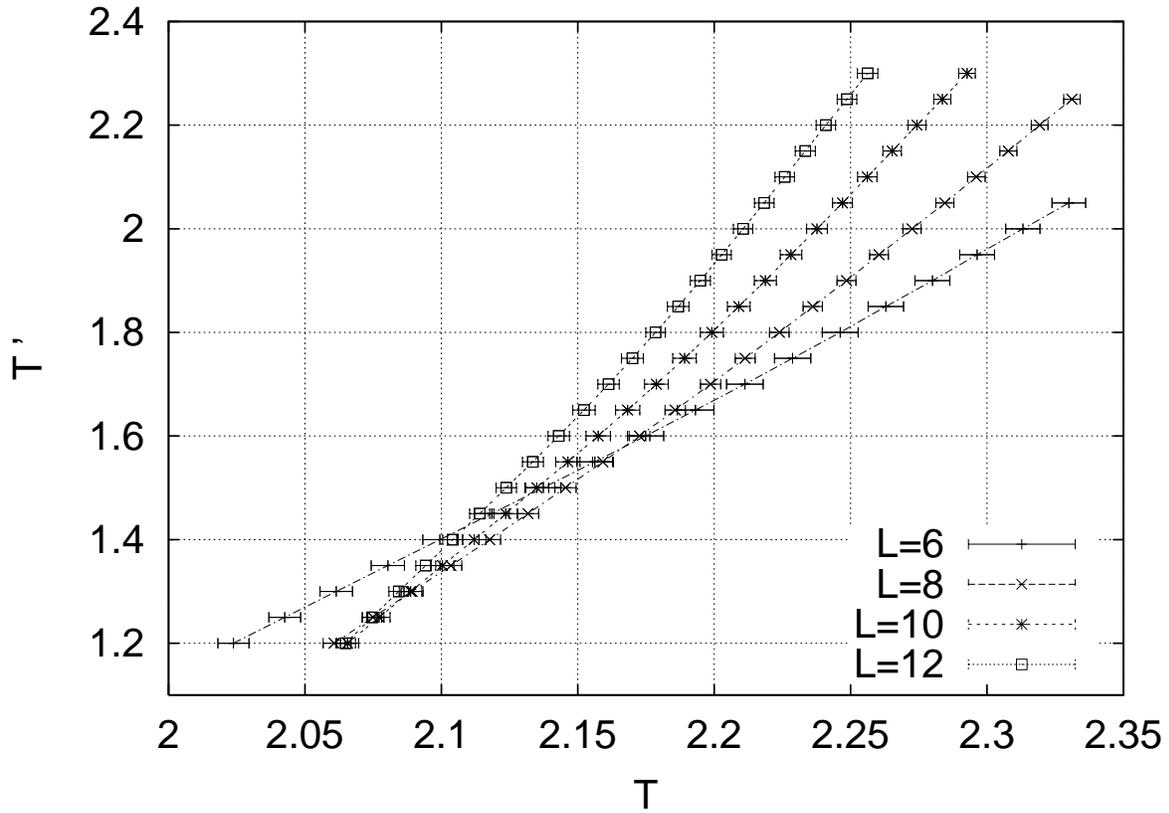}}}}
\caption{The renormalized temperature $T'$ as a function of the
temperature $T$.}
\label{fig:T}
\end{figure}

\begin{figure}
\resizebox{\textwidth}{!}{\rotatebox{270}{{\includegraphics{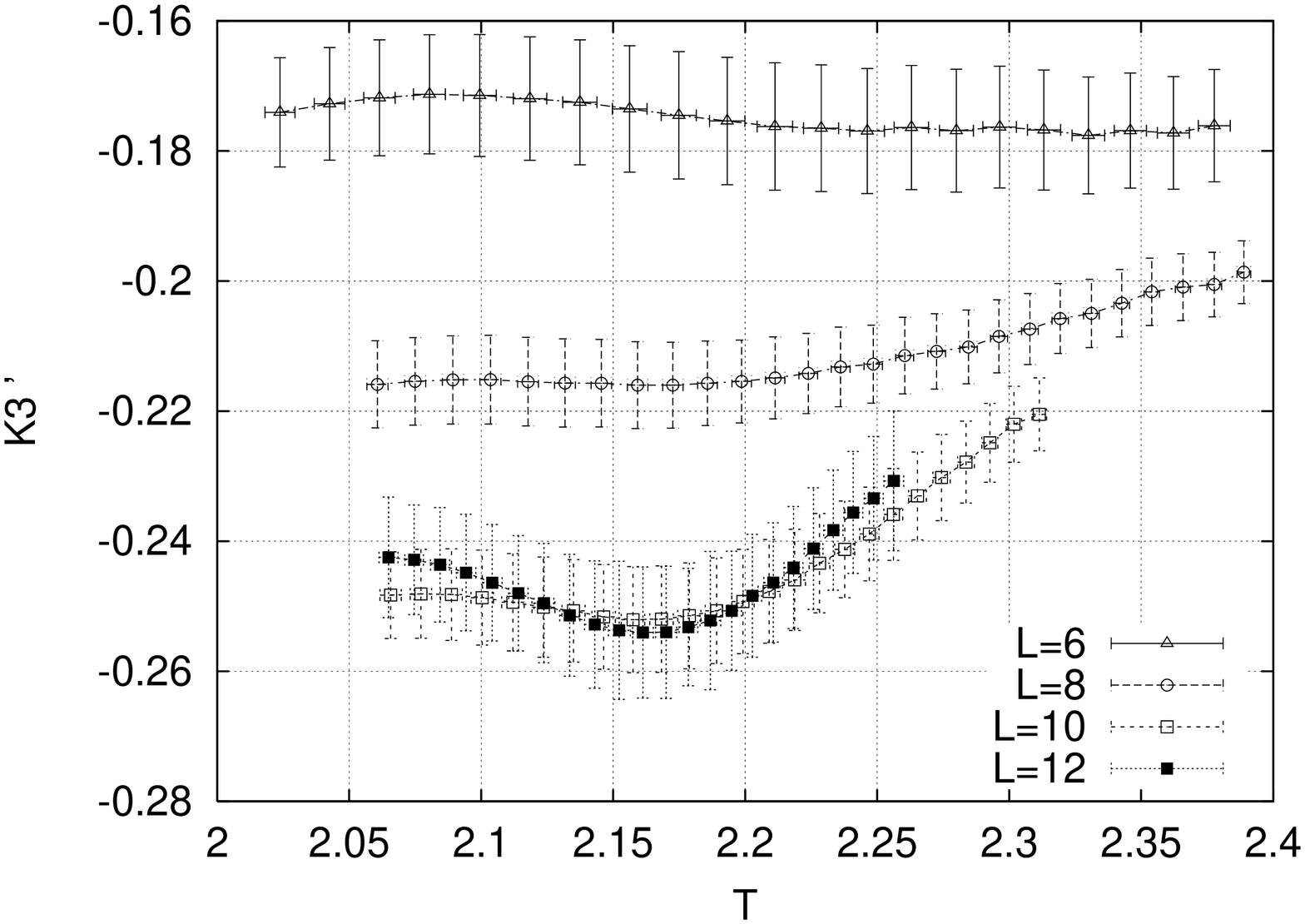}}}}
\caption{The renormalized coupling $K'_3$ as a function of the
temperature $T$. Negative values correspond to negative average
plaquette, \ie, increased frustration.}
\label{fig:K3}
\end{figure}

\begin{figure}
\resizebox{\textwidth}{!}{\rotatebox{270}{{\includegraphics{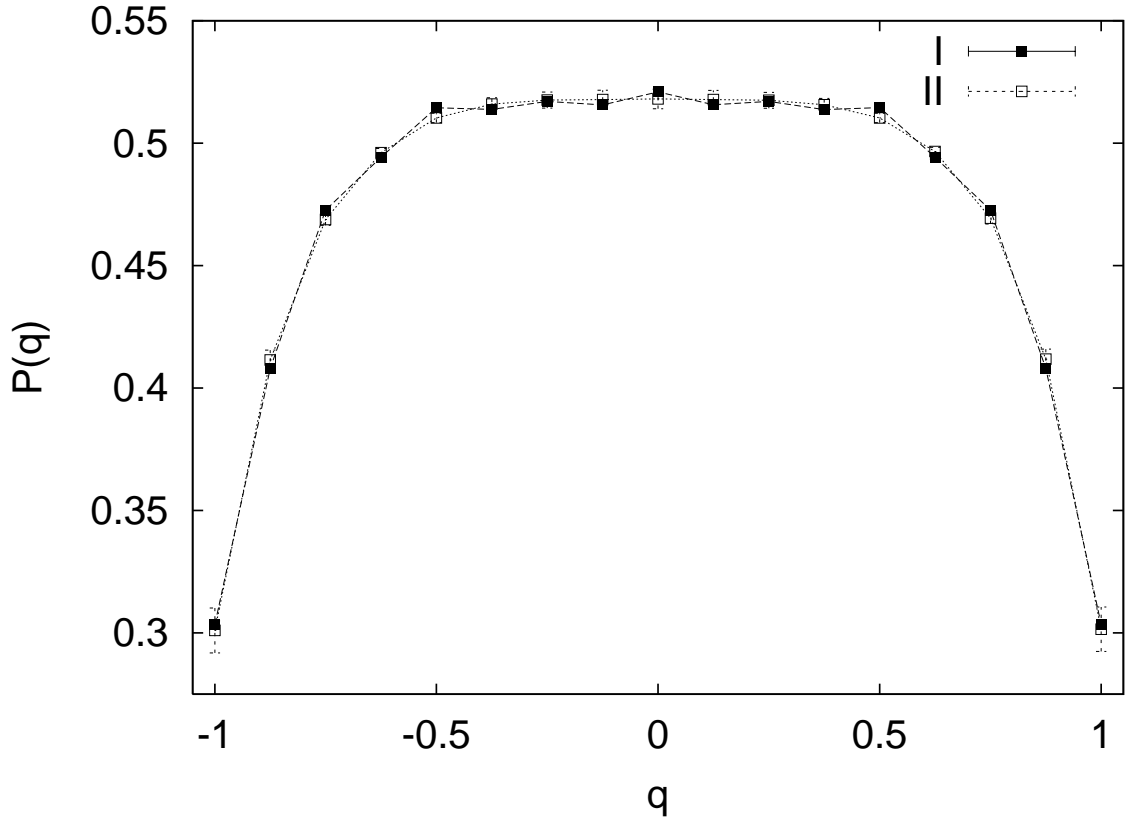}}}}
\caption{The overlap distribution of the $L=2$ system at
$T=1.3$, $K_3=-0.244(9)$ (I), and the renormalized overlap
distribution of the $L=12$ system at
$T=2.084(4)$ (II), corresponding to the matching point. }
\label{fig:match}
\end{figure}

The algorithm used in the MC runs is a parallel tempering (PT) Metropolis,
exploiting multi-spin coding. The fast computer code has been provided by
E.~Marinari and F.~Zuliani. 
Table~\ref{table:mc} lists the parameters used in the MC runs. 
For the purpose of studying the critical point it is unnecessary to measure
observables at very low temperature, and this considerably simplifies
the thermalization. Indeed, in the cold phase the renormalized
parameters flow rapidly to the zero-temperature fixed point.
The lowest temperature attained was about $0.93\,T_c$.  
Thermalization was verified by the symmetry of the $P^{(2)}_J(q)$ for
any given sample of $J$'s, while the efficiency of parallel tempering
was controlled through the histogram of the temperatures visited during
each run. In the MC runs on the $2^d$ system large statistics is
needed because of the large finite-size fluctuations of
both the spin glass and the pure gauge runs. 
Ten values of $K_3$ were considered, ranging from $-0.06$ to $-0.28$.  
The temperature spacing $\delta T$ is narrow in order to dispose of a dense
spanning of temperatures in the subsequent analysis.
\begin{table}
\begin{center}
\begin{tabular}{cccccccc}  \hline\hline
{\it L} & {\it Therm.} & {\it Measures} & {\it Samples} & $N_T$ &
$\delta T$ & $T_{min}$ & $T_{max}$ \\ \hline 
2& $2\times10^{5}$ & $3\times10^{5}$ & 32000 & 40 & 0.05 & 0.9 & 2.85 \\ 

6& $8 \times 10^{5}$ & $8 \times 10^{5}$ & 1024 & 21 & 0.025  &1.9 & 2.4 \\

8 & $10^{6}$ & $10^{6}$ & 1024 & 21 & 0.025  & 1.9 & 2.4 \\

10 & $10^{6}$ & $10^{6}$ & 1024 & 21 & 0.025  & 1.9 & 2.4 \\

12 & $10^{6}$ & $2 \times 10^{6}$ & 640 & 27 & 0.025  & 1.95 & 2.6 \\

\hline\hline
\end{tabular}
\end{center}
\caption{Parameters of the spin glass simulations. Columns correspond
to the linear size L, the number of MC-PT steps
for thermalization and measures, the number of $J$ samples, the number
of temperatures for the PT algorithm $N_T$, temperature step $\delta T$
and range $T_{min},\ T_{max}$.}
\label{table:mc}
\end{table}
\break
The overlap distributions of the small system, and the renormalized
overlap distributions of the large system
have been fitted with polynomials of $K_3$ and
$T$, respectively, and matched together for many values of $T'$.
This procedure avoids the necessity of bi-dimensional fits.
The matching was performed by numerical minimization of the $\chi^2$.
An example of matching of the overlap distributions is shown in
figure~\ref{fig:match}.    
Statistical errors on all the intermediate and final quantities were estimated
with jack-knife procedure on the data of the single $J$ samples.
The renormalized parameters $T'$ and $K'_3$ are plotted in
figures \ref{fig:T} and \ref{fig:K3}. 

\begin{figure}
\resizebox{\textwidth}{!}{\includegraphics{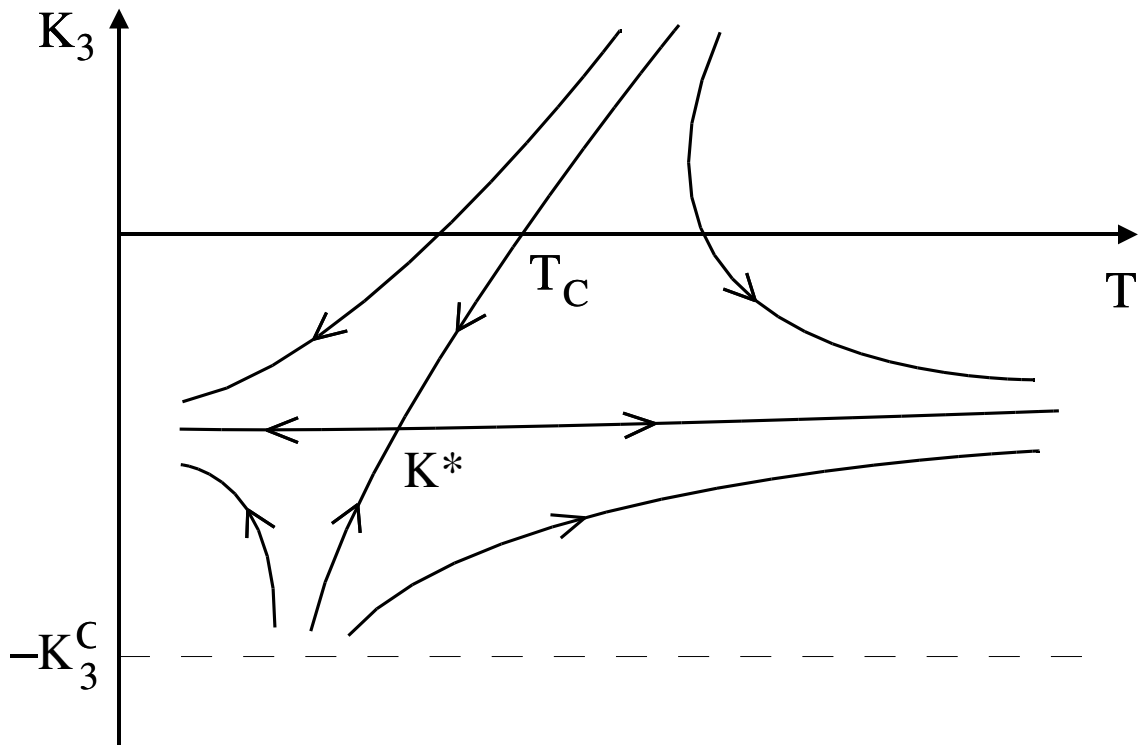}}
\caption{Qualitative picture of the RG flow showing the finite temperature
fixed point. The repulsive direction is approximately
horizontal; the critical temperature decreases for higher values of
$K_3$ because of the higher frustration.}
\label{fig:RGflow}
\end{figure}

The curves $T'(T;\,L)$ of figure~\ref{fig:T} cross in a same point,
within the error bars, apart from the curve for $L=6$ which is
affected by strong finite size effects. To obtain numerical estimates
of the crossing point and of the slopes, the curves have been
fitted, using the initial JK bins. The results are listed in
table~\ref{table:TcNu}. The curves $K_3(T;\, L)$ of
figure~\ref{fig:K3} are approximately constant close to the critical
temperature, and match for all temperatures for $L=10,\,12$. 
This means that in the plane $(T,\,K_3)$ the repulsive direction 
at the fixed point is orthogonal to the $K_3$ axis,
as sketched in figure~\ref{fig:RGflow}.
\begin{table}
\begin{center}
\begin{tabular}{ccc} \hline\hline
L/L' & $T_c$ & $\nu$ \\ \hline
8/6  & 2.17(2) & 0.87(6) \\
10/6 & 2.13(1) & 1.01(4) \\
12/6 & 2.109(9) & 1.05(3) \\
10/8 & 2.09(3) & 1.1(1) \\ 
12/8 & 2.07(1) & 1.07(6) \\
12/10 & 2.06(3) & 1.1(2) \\
\hline\hline
\end{tabular}
\end{center}
\caption{Estimates of the critical temperature $Tc$ and exponent $\nu$
obtained from the curves of figure~\ref{table:TcNu}.}
\label{table:TcNu}
\end{table}
Estimates of the critical temperature $Tc$ and of the exponent $\nu$
agree with previous numerical results by E.~Marinari and 
F.~Zuliani~\cite{MZ99},
obtained with the finite-size scaling of Binder cumulant, 
\ie $T_c= 2.03(3)$ and $\nu=1.0(1)$.

\section{Conclusions}

A renormalization group transformation suitable for spin glasses, and more
generally for disordered systems, has been introduced for the first time. 
The procedure is non-standard for two reasons. 1.~The coarse graining
transformation is performed on the distribution of site overlaps,
which clearly is not a Gibbs--Boltzmann distribution. 2.~The space of such
distributions is parameterized by additional terms in the distribution
of the disorder, while the Hamiltonian of the system is kept fixed.
This leads to a quantitative definition of universality classes 
for spin glass models, clarifying also the connections between $\Z_2$
and Gaussian spin glasses, the Ising model, the fully frustrated
model. 

This RG scheme has been tested numerically with Monte Carlo method in
the case of the 4-dimensional EA spin glass. Good estimates of the
critical temperature and of the exponent $\nu$ have been obtained with
moderate computer time.

A plot of the RG flow in spin glass models has been sketched for the first
time.

\section{Acknowledgements} 
The authors wish to thank E. Marinari and F. Zuliani for providing a
computer code for the SG Monte Carlo and for interesting discussions.


\end{document}